\documentclass[conference]{IEEEtran}

\usepackage{graphicx}
\usepackage{amsmath,bbm,epsfig,amssymb,amsfonts, amstext, verbatim,amsopn,cite,subfigure,multirow,multicol,lipsum}
\usepackage{balance}
\usepackage{url}
\usepackage{amsfonts}
\usepackage{epsfig}
\usepackage{setspace}
\usepackage{stmaryrd}
\usepackage{psfrag}	
\usepackage{multirow}
\usepackage{float}
\usepackage[process=auto]{pstool}
\usepackage{etoolbox}
\usepackage{algorithm}
\usepackage{algorithmic}
\usepackage{hyperref}

\usepackage{pifont}
\allowdisplaybreaks

\newtheorem{theo}{Theorem}
\newtheorem{lem}{Lemma}
\newtheorem{remk}{Remark}

\newtoggle{ConfVersion}

\togglefalse{ConfVersion}

\iftoggle{ConfVersion}{%
}{%
}

\EndPreamble
\begin{document}

\title{Statistical Modeling of FSO Fronthaul Channel \\ for Drone-based Networks
\vspace{-0.3cm}}
\author{Marzieh Najafi$^{\dag}$, Hedieh Ajam$^{\dag}$, Vahid Jamali$^{\dag}$,  Panagiotis D. Diamantoulakis$^{\ddag}$,\\ George K. Karagiannidis$^{\ddag}$, and Robert Schober$^{\dag}$\\  
$^{\dag}$Friedrich-Alexander University of Erlangen-Nuremberg, Germany \\
$^{\ddag}$Aristotle University of Thessaloniki, Greece
\vspace{-0.2cm}}

\maketitle
\begin{abstract}
We consider a drone-based communication network, where several drones hover above an area and serve as mobile remote radio heads for a large number of mobile users. We assume that the drones employ free space optical (FSO) links for fronthauling of the users' data to a central unit. The main focus of this paper is to quantify the geometric loss of the FSO channel arising from random fluctuation of the position and orientation of the drones. In particular, we derive upper and lower bounds, corresponding approximate expressions, and a closed-form statistical model for the geometric loss. Simulation results validate our derivations and quantify the FSO channel quality as a function of the drone's instability, i.e., the  variation of its position and orientation.
\end{abstract}


\section{Introduction}
Recently, there has been a growing interest in unmanned aerial vehicles (UAVs) and drones for civil applications, such as delivering cellular and internet services to remote regions or areas where a large number of users is \textit{temporarily} gathered, e.g., a football match or a live concert, where \textit{permanent infrastructure} does not exist or is costly to deploy \cite{Alouini_Drone,Data_Collection_UAV_FSO}. 
In particular, drones may hover above the desired area and operate as mobile remote radio heads to assist the communication between the users and a central unit (CU) \cite{Alouini_Drone}. 

For these applications, free space optical (FSO) systems have been considered as promising candidates for fronthauling of the data gathered by the drones to the CU \cite{Alouini_Drone}. FSO systems offer the large bandwidths needed for data fronthauling and FSO transceivers are cheap and easy to implement \cite{Optic_Tbit,MyTCOM}. However, the main factor that deteriorates the quality of the FSO link between a hovering drone and the CU is the instability of the drone, i.e., the variation of its position and orientation. Therefore, an immediate question is: \textit{How good does the drone have to be in maintaining its position and orientation in order to achieve a certain FSO link quality?} The goal of this paper is to answer this question for uplink transmission by characterizing the geometric loss\footnote{\label{Ftn:Loss} The receiver can only capture that fraction of  power that falls onto the area of its  photo-detector. This phenomenon is known as geometric loss. On the other hand, pointing errors further increase the geometric loss. This phenomenon is also known as misalignment loss \cite{Steve_pointing_error}. For simplicity of presentation,  in this paper, we refer to the combined effect of these impairments as geometric loss. } caused by random fluctuations of the drone's position and orientation. 

We note that,  even in conventional FSO links where both transceivers are mounted on top of buildings, random fluctuations of the transceivers' positions occur due to building sway which leads to a random geometric loss. For this case, corresponding statistical models were developed in \cite{Steve_pointing_error} and \cite{Alouini_Pointing}. However, the geometric loss for the case when the transmitter is a drone requires a new statistical model due to following differences:
\textit{i)} Unlike building sway, where the buildings exhibit limited movement due to wind loads and thermal expansion, for drone-based FSO communication, both the position and the orientation of the drone may fluctuate over time and have to be modelled as random variables (RVs).
\textit{ii)} For conventional FSO links, it is assumed that the laser beam is orthogonal with respect to (w.r.t.) the photo-detector (PD) plane at the receiver. However, this assumption may not hold for drone-based FSO communication. For example, the PD at the CU may receive data from several drones having different positions. Hence, it is not possible that the laser beams of all drones are orthogonal to the PD plane. Also, the positions of the drones may change due to changing traffic needs and the CU may not be able to adapt the orientation of the PD due to limited mechanical capabilities.

Drones and UAVs with FSO links have already been considered in the literature \cite{Alouini_Drone,UAV_FSO,Data_Collection_UAV_FSO,Inter-UAV_FSO}. In particular, \cite{Alouini_Drone} discussed the advantages and challenges of FSO fronthauling for drone-based networks. Moreover, \cite{UAV_FSO,Data_Collection_UAV_FSO,Inter-UAV_FSO} studied a system consisting of several drones that communicate with each other through FSO links. Specifically, the authors of \cite{UAV_FSO} focused on the derivation of a \textit{deterministic} model for the geometric loss assuming that the laser beam is always \textit{orthogonal} to the receiver's PD plane. However, to the best of the authors' knowledge, a statistical model for the geometric loss of  drone-based FSO links, which takes into account the fluctuations of the drone's position and orientation as well as the non-orthogonality of the laser beam w.r.t. the PD plane, has not been developed in the literature, yet.

In this paper, we model the geometric loss of the drone-based FSO fronthaul channel while taking into account the drone's instability and the non-orthogonality of the laser beam w.r.t. the PD plane. To this end, we first model the position and orientation of the drone as RVs. Then, we derive the geometric loss for a given realization of these RVs. In particular, we derive upper and lower bounds as well as approximate expressions for the geometric loss which are simpler than the exact expression. Finally, we derive a statistical model for the geometric loss assuming the drone's position and orientation follow Gaussian distributions. Our simulation results validate the accuracy of the proposed bounds, approximations, and statistical model and quantify the quality of the FSO channel as a function of the drone's instability.

\section{System and Channel Models}
In this section, we present the considered system model and the FSO channel model. 
\subsection{System Model}

\begin{figure}[t]
\centering
\scalebox{0.9}{
\pstool[width=1\linewidth]{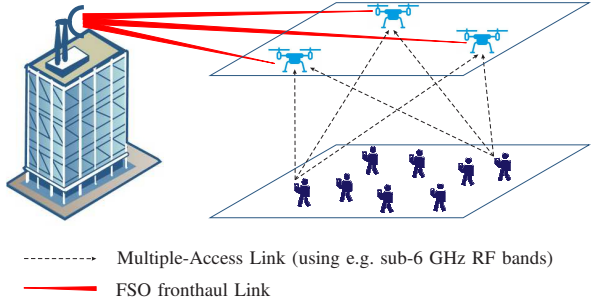}{
\psfrag{RF}[l][c][0.8]{Multiple-Access Link (using e.g. sub-6 GHz RF bands)}
\psfrag{FS}[l][c][0.8]{FSO fronthaul Link}
}}
\caption{Proposed drone-based communication system where the drones communicate with the mobile users via an RF multiple-access link and with the central unit via FSO fronthaul links.}
\label{Fig:SysMod}\vspace{-0.3cm}
\end{figure}

We consider a drone-based communication network, where an arbitrary number of drones hover above an area, where a large number of users are concentrated, and operate as mobile remote radio heads to assist the communication between the users and a CU, see Fig.~\ref{Fig:SysMod}. In particular, we consider an uplink scenario where mobile users send their data to the drones over a multiple-access link, e.g., using sub-6 GHz radio frequency (RF) bands, and the drones forward the data over  fronthaul links to a CU for final processing. Forwarding the aggregated user data received at the drones to the CU requires a huge data rate for the fronthaul links. Hereby, we propose to establish FSO links between the drones and the CU as large bandwidths can be realized at optical frequencies.  The main goal of this paper is to develop a mathematical model that captures the effect of the fluctuation of a hovering drone's position and orientation on the FSO channel quality. To do so,  we formally define the position and orientation of the drone and the CU in our system model in the following. 

In order to characterize an object in three dimensions, we need \textit{at most} six independent variables, namely three variables to specify the position of a reference point of the object and three variables to quantify its orientation.  With this in mind, we characterize the position and orientation of the CU and the drone as follows.

\subsubsection{CU} The CU is a fixed node located on top of a building. Without loss of generality, we choose the center of the PD as the reference point, which is located in the origin of the Cartesian coordinate system $(x,y,z)=(0,0,0)$. This coordinate system is referred to as Coordinate System~1,  cf. Fig.~\ref{Fig:Nodes}. Moreover, we assume a circular PD of radius $a$. Note that it suffices to characterize the plane in which the PD lies to specify its orientation. Here, without loss of generality, we assume the PD lies in the $y-z$ plane at $x=0$.

\subsubsection{Drone} 
\begin{figure}[t]
\centering
\scalebox{0.7}{
\pstool[width=1\linewidth]{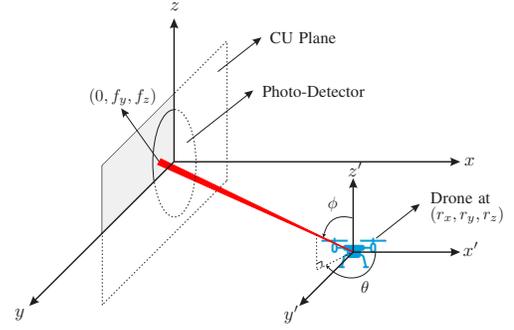}{
\psfrag{X}[c][c][1]{$x$}
\psfrag{Y}[c][c][1]{$y$}
\psfrag{Z}[c][c][1]{$z$}
\psfrag{A}[c][c][1]{$x'$}
\psfrag{B}[c][c][1]{$y'$}
\psfrag{C}[c][c][1]{$z'$}
\psfrag{S}[l][c][0.9]{CU Plane}
\psfrag{F}[l][c][0.9]{Photo-Detector}
\psfrag{D1}[l][c][0.9]{Drone at}
\psfrag{D2}[l][c][0.9]{$(r_x,r_y,r_z)$}
\psfrag{R}[c][c][0.9]{$(0,f_y,f_z)$}
\psfrag{P}[c][c][0.9]{$\phi$}
\psfrag{T}[c][c][0.9]{$\theta$}
}}
\caption{The position and orientation of the CU and the drone in the considered coordinate systems.}
\label{Fig:Nodes}\vspace{-0.3cm}
\end{figure}

For the communication system under consideration, the parameters that directly affect the FSO channel are the position of the laser source of the drone and the direction of the laser beam. Therefore, without loss of generality, we refer to the position of the laser source  and the direction of the laser beam as the drone's position and orientation, respectively.   We assume that the drone is in the hovering state. For practical reasons, in the hovering state, the position and orientation of the drone cannot be perfectly constant. Therefore, we model them as RVs.  In particular, let $\mathbf{r}=(r_x,r_y,r_z)$ denote the vector of random position variables of the drone. 
Furthermore, let  $\boldsymbol{\omega} = (\theta,\phi,\gamma)$ denote the vector of random orientation variables. The value of $\boldsymbol{\omega}$ depends on the coordinate system which is used to quantify these variables. Without loss of generality, we hereby use the following coordinate system which simplifies our analysis. For a given  $\mathbf{r}$, let us define a new coordinate system, denoted as Coordinate System~2, with $\mathbf{r}$ as its origin and axes  $x'$, $y'$, and $z'$, which are parallel to the $x$, $y$, and $z$ axes, respectively, cf. Fig.~\ref{Fig:Nodes}. We use variables $\theta$ and $\phi$ to determine the direction of the laser beam in a spherical representation of Coordinate System~2. In particular, $\theta\in[0,2\pi]$ denotes the angle between axis $x'$ and the projection of the beam vector onto the $x'-y'$ plane and $\phi\in[0,\pi]$ represents the angle between the beam vector and the $z'$ axis. The third orientation variable $\gamma$ is used to quantify the rotation around the beam vector. The advantage of the aforementioned representation of the orientation variables is two-fold. First, variable  $\boldsymbol{\omega}$ does not change if coordinate $\mathbf{r}$ changes, i.e., the position and orientation variables are independent. Second, a rotation around the beam line does not affect the signal at the PD assuming rotational beam symmetry. Therefore, the value of $\gamma$ is irrelevant for our analysis.  Hence, for simplicity of presentation, we drop  $\gamma$ and use  $\boldsymbol{\omega} = (\theta,\phi)$  in the remainder of the paper. 

\subsection{FSO Channel Model}
We assume direct detection at the CU where the PD responds to changes in the received optical signal  power \cite{FSO_Survey_Murat}. 
 In general, the FSO channel coefficient, denoted by $h$, is affected by several factors and can be modelled as follows 
\begin{IEEEeqnarray}{lll}\label{Eq:channel}
h=\eta h_p h_a h_g,
\end{IEEEeqnarray}
where $\eta$ is the responsivity of the PD and $h_p$, $h_a$, and $h_g$ are the path loss,  atmospheric turbulence loss,  and geometric loss, respectively. In particular, path loss $h_p$ is deterministic and represents the power loss over a propagation path due to attenuation. Atmospheric turbulence loss $h_a$ is random and induced 
by inhomogeneities in the temperature and the pressure of the atmosphere and is typically modelled as log-normal or Gamma-Gamma 
distributed RV. Moreover, geometric loss $h_g$ is caused by the divergence of the optical beam between the transmitter and the PD and the misalignment of the laser beam line and the center of the PD  \cite{FSO_Survey_Murat,FSO_Vahid}. Fluctuations of the drone's position and orientation lead to a random geometric loss $h_g$. Hence, in this paper, we develop a statistical model for the geometric loss. This model  allows us to study the performance of the considered communication system and the impact of the system parameters, such as the ability of the drone to maintain its position and orientation, on the FSO channel quality.  


\section{Modeling of the Geometric Loss}
 In this section, we first derive a deterministic model for the geometric loss for a given position and orientation of the nodes, and then a statistical model, assuming that the drone's position and orientation fluctuate in the hovering state.

\subsection{Deterministic Model}\label{Sec:hg_det}
Here, we derive the geometric loss for a given state of the drone, i.e., for given $\mathbf{r}$ and $\boldsymbol{\omega}$. To do so, we first find the center of the beam footprint and the power density on the PD plane. Using these results, we then derive the geometric loss.

\subsubsection{Center of Beam Footprint}

The line of the beam can be represented in Cartesian Coordinate System~1 as follows
\begin{IEEEeqnarray}{lll} \label{Eq:BeamLine}
(x,y,z) = (r_x,r_y,r_z) + t (d_x,d_y,d_z),
\end{IEEEeqnarray}
where $t$ is an arbitrary real number and $\mathbf{d}=(d_x,d_y,d_z)$ denotes the beam direction which can be found as a function of $\theta$ and $\phi$ as 
\begin{IEEEeqnarray}{lll} \label{Eq:d_angle}
\mathbf{d}=\big(\sin(\phi)\cos(\theta),\sin(\phi)\sin(\theta),\cos(\phi)\big).
\end{IEEEeqnarray}

The center of the beam footprint on the PD can be obtained as the intersection point of the line of the laser beam and the PD plane, $x=0$. Denoting   the center of the footprint of the beam on the PD as $\mathbf{f}=(f_x,f_y,f_z)$, cf. Fig.~\ref{Fig:Nodes}, we obtain
\begin{IEEEeqnarray}{lll} \label{Eq:FootPrint_Center}
\mathbf{f} = \left(0,r_y-r_x\tan(\theta),r_z-r_x\frac{\cot(\phi)}{\cos(\theta)}\right).
\end{IEEEeqnarray}

\subsubsection{Power Density on the PD Plane}

We assume a Gaussian beam which dictates that the power distribution in any plane perpendicular to the direction of the wave propagation follows a Gaussian distribution \cite{FSO_Survey_Murat}. In particular, let us consider a perpendicular plane where the distance between the center of the beam footprint on the plane and the laser source is denoted by $L$. Then, the power density for any point on this perpendicular plane with distance $l$ to the center of the beam footprint is given by \cite{Steve_pointing_error}
\begin{IEEEeqnarray}{lll} \label{Eq:PowerOrthogonal}
I^{\mathrm{orth}}(L,l) = \frac{2}{\pi w^2(L)}\exp\left(-\frac{2l^2}{w^2(L)}\right),
\end{IEEEeqnarray}
where $w(L)$ is the beam width and is obtained as
\begin{IEEEeqnarray}{lll} \label{Eq:BeamWidth}
w(L) = w_0\sqrt{1+\left(1+\frac{2w_0^2}{\rho^2(L)}\right)\left(\frac{\lambda L}{\pi w_0^2}\right)^2}.
\end{IEEEeqnarray}
In (\ref{Eq:BeamWidth}), $w_0$ denotes beam waist radius and $\rho(L)=(0.55C_n^2k^2L)^{-3/5}$ is the coherence length, where $C_n^2$ is the index of refraction structure parameter (assumed to be constant along the propagation path), $k = 2\pi/\lambda$ is the optical wave-number, and $\lambda$ is the optical wavelength.
As mentioned before, for the problem at hand, the plane of the PD is not necessarily orthogonal to the beam direction. For this case, the power density in the PD plane, denoted by $I(y,z)$, is given in the following lemma.

\begin{lem}\label{Lem:PowerDensity}
Under the mild conditions $\Vert \mathbf{r} \Vert \gg \Vert \mathbf{f} \Vert$ and $\Vert \mathbf{r} \Vert \gg   \Vert (y,z) \Vert$, where $\|\cdot\|$  denotes the norm of a vector, the power density at point $(y,z)$ on the PD plane  is given by
\begin{IEEEeqnarray}{lll} \label{Eq:Intensity}
I(y,z)&=\sin(\psi)I^{\mathrm{orth}}\big(L(\mathbf{r}),  l(\boldsymbol{\omega},(y,z))\big) \nonumber\\
&=\frac{2\sin (\psi) }{\pi w^2(L)}\exp\Big(\frac{-2}{ w^2(L)}(\rho_{y}\tilde{y}^2+\rho_{z}\tilde{z}^2
+ 2\rho_{yz}\tilde{y}\tilde{z})\Big),\quad\,\,\,\,
\end{IEEEeqnarray}
where $\psi=\sin^{-1}(\sin(\phi) \cos(\theta))$, $L(\mathbf{r})=\Vert \mathbf{r} \Vert$, $l(\boldsymbol{\omega},(y,z))=\rho_{y}\tilde{y}^2+\rho_{z}\tilde{z}^2
+ 2\rho_{yz}\tilde{y}\tilde{z}$, $\tilde{y} = y-f_y$, $\tilde{z} =z-f_z$, and $I^{\mathrm{orth}}(\cdot,\cdot)$ is given by (\ref{Eq:PowerOrthogonal}). Moreover, $\rho_{y} = \cos^2(\phi)+\sin^2(\phi)\cos^2(\theta)$,
$\rho_{z} = \sin^2(\phi)$, and
$\rho_{yz} = -\cos(\phi)\sin(\phi)\sin(\theta)$.
\end{lem}
\begin{IEEEproof}
Please refer to Appendix~\ref{App:Lem_PowerDensity}.
\end{IEEEproof}

Note that the conditions under which (\ref{Eq:Intensity}) in Lemma~\ref{Lem:PowerDensity} holds are met in practice, as in typical FSO links, $\Vert \mathbf{r} \Vert$ is on the order of several hundred meters, whereas $\Vert \mathbf{f} \Vert$ and $\Vert (y,z) \Vert$ are on the order of a  few  centimeters.

\subsubsection{Geometric Loss}

The fraction of power collected at the PD, denoted by $h_g(\mathbf{r},\boldsymbol{\omega})$, can be obtained by integrating the power density obtained in Lemma~\ref{Lem:PowerDensity} over the PD area. This leads to
\begin{IEEEeqnarray}{lll}\label{Eq:PowerPhotoDetector} 
h_g(\mathbf{r},\boldsymbol{\omega}) =  \underset{(y,z)\in\mathcal{A}}{\iint} 
I(y,z) 
\mathrm{d}y\mathrm{d}z,
\end{IEEEeqnarray}
where $I(y,z) $ is given in (\ref{Eq:Intensity}) and $\mathcal{A}$ is the set of $(y,z)$ within the PD area, i.e., $\mathcal{A}=\{(y,z)|y^2+z^2\leq a^2\}$. Unfortunately, the exact value of $h_g(\mathbf{r},\boldsymbol{\omega})$  cannot be derived in closed form. Instead, in the following theorem, we provide an upper and a lower bound on $h_g(\mathbf{r},\boldsymbol{\omega})$ which are subsequently used to derive an approximate closed-form expression for $h_g(\mathbf{r},\boldsymbol{\omega})$. 

\begin{theo}\label{Theo:Power}
Using Lemma~\ref{Lem:PowerDensity}, the geometric loss $h_g(\mathbf{r},\boldsymbol{\omega})$ is lower bounded by $h_g^{\mathrm{low}}(\mathbf{r},\boldsymbol{\omega})$ and upper bounded by $h_g^{\mathrm{upp}}(\mathbf{r},\boldsymbol{\omega})$ where
\begin{IEEEeqnarray}{lll} \label{Eq:upper_lower_bound}
 h_g^{\mathrm{low}}(\mathbf{r},\boldsymbol{\omega}) =\frac{2\sin (\psi) }{\pi w^2(L)}\times \nonumber\\
 \underset{(y,z)\in\mathcal{A}}{\iint} 
\exp\left(-\frac{2}{ w^2(L)}\Big(\frac{1}{\rho_{\min}}(y-u)^2+\frac{1}{\rho_{\max}}{z}^2\Big)\right) 
\mathrm{d}y\mathrm{d}z\IEEEyesnumber\IEEEyessubnumber\qquad\\
 h_g^{\mathrm{upp}}(\mathbf{r},\boldsymbol{\omega}) =\frac{2\sin (\psi) }{\pi w^2(L)}\times\nonumber\\
 \underset{(y,z)\in\mathcal{A}}{\iint} 
\exp\left(-\frac{2}{ w^2(L)}\Big(\frac{1}{\rho_{\max}}(y-u)^2+\frac{1}{\rho_{\min}}{z}^2\Big)\right)
\mathrm{d}y\mathrm{d}z. \,\, \IEEEyessubnumber\quad
\end{IEEEeqnarray}
Here, $u=\sqrt{f_y^2+f_z^2}$, $\rho_{\min}=\frac{2}{\rho_y+\rho_z+\sqrt{(\rho_y-\rho_z)^2+4\rho_{zy}^2}}$, and $\rho_{\max}=\frac{2}{\rho_y+\rho_z-\sqrt{(\rho_y-\rho_z)^2+4\rho_{zy}^2}}$.  For the special case, where the beam line is orthogonal to the plane of the PD, we obtain $\rho_{\max}=\rho_{\min}=1$ and $\rho_{yz}=0$ and the upper and lower bounds coincide and become identical to $h_g(\mathbf{r},\boldsymbol{\omega})$.
\end{theo}
\begin{IEEEproof}
Please refer to Appendix~\ref{App:Integral}.
\end{IEEEproof}

\begin{figure}[t]
\centering
\scalebox{0.7}{
\pstool[width=1\linewidth]{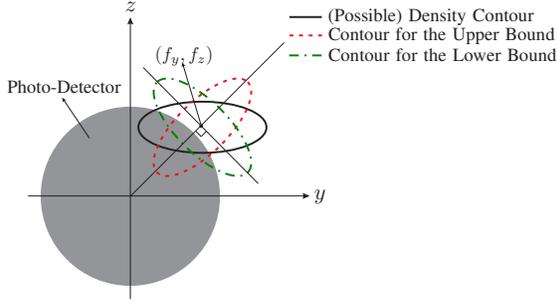}{
\psfrag{Y}[c][c][1.2]{$y$}
\psfrag{Z}[c][c][1.2]{$z$}
\psfrag{D}[c][c][1]{Photo-Detector}
\psfrag{F}[c][c][1]{$(f_y,f_z)$}
\psfrag{A}[l][c][1]{(Possible) Density Contour}
\psfrag{B}[l][c][1]{Contour for the Upper Bound}
\psfrag{C}[l][c][1]{Contour for the Lower Bound}
}}
\caption{A possible power density contour in the PD plane and the contours used to derive the upper and lower bounds.}\vspace{-0.5cm}
\label{Fig:Contour}
\end{figure}

\begin{remk}
We use Fig.~\ref{Fig:Contour} to illustrate the basic idea behind the  upper and lower bounds proposed in Theorem~\ref{Theo:Power}.  In particular, unlike the case where the optical beam is orthogonal w.r.t. the PD plane and the power density contours are circles \cite{Steve_pointing_error}, the case where the optical beam is non-orthogonal w.r.t. the PD plane leads to power density contours which are \textit{rotated ellipses}, e.g., the black solid contour in Fig.~\ref{Fig:Contour}. We have derived the lower bound assuming a contour that is a rotated ellipse whose major axis is perpendicular to the line connecting the center of the footprint to the origin, e.g., the green dash-dotted contour in Fig.~\ref{Fig:Contour}. Moreover, for the upper bound, the contour is a rotated ellipse whose minor axis is perpendicular to the line connecting the center of the footprint to the origin, e.g., the red dotted contour in Fig.~\ref{Fig:Contour}. In the special case where the major (minor) axis of the \textit{original} power density contour is perpendicular to the line connecting the center of the footprint to the origin, the upper (lower) bound matches the exact geometric loss.
\end{remk}

We emphasize that even for the case when the beam line is orthogonal w.r.t. the PD plane, the exact value of $h_g(\mathbf{r},\boldsymbol{\omega})$ is cumbersome and provides little insight. Therefore, in \cite{Steve_pointing_error}, the authors proposed an approximation which was shown to be very accurate for $w(L)/a\geq 6$ and has been widely used subsequently \cite{Steve_MISO_FSO,FSO_Receivers_UAV,George_Pointing_error,Alouini_Pointing_Beckman_Hoyt}. The proposed bounds in Theorem~\ref{Theo:Power} have two main advantages. First, for the special case when the beam line is orthogonal to the PD plane, the upper and lower bounds coincide with the exact  $h_g(\mathbf{r},\boldsymbol{\omega})$. Second, the form of the integrals in (\ref{Eq:upper_lower_bound}) allows us to employ the same technique as in \cite[Appendix]{Steve_pointing_error} to obtain accurate approximations. In particular, $h_g^{\mathrm{low}}(\mathbf{r},\boldsymbol{\omega})$ and $h_g^{\mathrm{upp}}(\mathbf{r},\boldsymbol{\omega})$ in (\ref{Eq:upper_lower_bound}) can be approximated by \cite{Steve_pointing_error}
\begin{IEEEeqnarray}{lll} \label{Eq:upper_lower_cal}
\tilde{h}_g^{\mathrm{low}}(\mathbf{r},\boldsymbol{\omega}) = A_0 \exp\left(-\frac{2u^2}{k_{\min}w^2(L)}\right)  \IEEEyesnumber\IEEEyessubnumber\quad\\
 \tilde{h}_g^{\mathrm{upp}}(\mathbf{r},\boldsymbol{\omega}) = A_0 \exp\left(-\frac{2u^2}{k_{\max}w^2(L)}\right),\IEEEyessubnumber\quad
\end{IEEEeqnarray}
respectively, where $A_0=\mathrm{erf}(\nu_{\min})\mathrm{erf}(\nu_{\max})$, $k_{{\min}}=\frac{\sqrt{\pi}\rho_{\min}\mathrm{erf}(\nu_{\min})}{2\nu_{\min}\exp(-\nu_{\min}^2)}$, and $\nu_{\min}=\frac{a}{w(L)}\sqrt{\frac{\pi}{2\rho_{\min}}}$. Similarly, $k_{{\max}}$ and  $\nu_{\max}$ are obtained by changing the index $\min$ to $\max$ in the relevant equations. Moreover, $\mathrm{erf}(x)=\frac{1}{\sqrt{\pi}}\int_{-x}^{x}\exp(-t^2)\mathrm{d}t$ is the error function. Note that the only difference between the approximate upper and lower bounds in (\ref{Eq:upper_lower_cal}) are the terms $k_{\max}$ and $k_{\min}$, respectively. This motivates us to propose the following  approximation of $h_g(\mathbf{r},\boldsymbol{\omega})$
\begin{IEEEeqnarray}{lll} \label{Eq:hg_mean}
\tilde{h}_g(\mathbf{r},\boldsymbol{\omega}) = A_0 \exp\left(-\frac{2u^2}{k_{\mathrm{mean}}w^2(L)}\right),
\end{IEEEeqnarray}
where $k_{\mathrm{mean}}=\frac{k_{\min}+k_{\max}}{2}$. Therefore, instead of considering the approximate upper and lower bounds in (\ref{Eq:upper_lower_cal}), in the following, we employ the approximation in (\ref{Eq:hg_mean}) for our statistical analysis. We show in Section \ref{Sec:Sim} that this approximation is accurate for a range of simulation parameters.

\subsection{Statistical Model}

In the previous subsection, we derived an approximation of the geometric loss $h_g(\mathbf{r},\boldsymbol{\omega})$ for a drone's given state $\mathbf{r}$ and $\boldsymbol{\omega}$ in (\ref{Eq:hg_mean}). However, in practice, the position and orientation of a hovering drone fluctuates, and hence, $\mathbf{r}$ and $\boldsymbol{\omega}$ are RVs. In the following, we first discuss the means and variances of these RVs.

\subsubsection{Mean} Let $\boldsymbol{\mu}_{\mathbf{r}}=(\mu_x,\mu_y,\mu_z)$ and $\boldsymbol{\mu}_{\boldsymbol{\omega}}=(\mu_{\theta},\mu_{\phi})$ denote the means of RVs $\mathbf{r}$ and $\boldsymbol{\omega}$, respectively. Since the drone is supposed to hover above the users, the mean position $\boldsymbol{\mu}_{\mathbf{r}}$  depends on the location of the users as well as the desired operating height of the drone. Given $\boldsymbol{\mu}_{\mathbf{r}}$, the drone's tracking system aims to determine  $\boldsymbol{\mu}_{\boldsymbol{\omega}}$ such that the beam line intersects with the center of the PD, i.e., $(0,0,0)$. This leads to
\begin{IEEEeqnarray}{lll} \label{Eq:Angles}
\mu_{\theta}=\begin{cases}
\tan^{-1}\left(\frac{\mu_y}{\mu_x}\right)&\mathrm{if}\,\,\mu_x>0 \\
\pi+\tan^{-1}\left(\frac{\mu_y}{\mu_x}\right)&\mathrm{otherwise}
\end{cases} 
\IEEEyesnumber\IEEEyessubnumber \\
\mu_{\phi}=\pi-\cos^{-1}\Bigg(\frac{\mu_z}{\sqrt{\mu_x^2+\mu_y^2+\mu_z^2}}\Bigg).
\IEEEyessubnumber
\end{IEEEeqnarray}
 Note that  the values of $\mu_{\theta}$ and $\mu_{\phi}$ may deviate from the above results if there is a tracking error. Nevertheless, in this paper, we assume perfect tracking where for a given $\boldsymbol{\mu}_{\mathbf{r}}$, $\boldsymbol{\mu}_{\boldsymbol{\omega}}$ is  obtained from (\ref{Eq:Angles}).

\subsubsection{Variance} Let $\boldsymbol{\sigma}_{\mathbf{r}}=(\sigma_x,\sigma_y,\sigma_z)$ and $\boldsymbol{\sigma}_{\boldsymbol{\omega}}=(\sigma_{\theta},\sigma_{\phi})$ denote the standard deviations of RVs $\mathbf{r}$ and $\boldsymbol{\omega}$, respectively. The values of $\boldsymbol{\sigma}_{\mathbf{r}}$ and $\boldsymbol{\sigma}_{\boldsymbol{\omega}}$ depend on how well the drone is able to maintain its position and orientation around the mean values $\boldsymbol{\mu}_{\mathbf{r}}$ and  $\boldsymbol{\mu}_{\boldsymbol{\omega}}$, respectively. The smaller the values of the elements of $\boldsymbol{\sigma}_{\mathbf{r}}$ and $\boldsymbol{\sigma}_{\boldsymbol{\omega}}$ are, the more stable the drone is. Hence, we consider $\boldsymbol{\sigma}_{\mathbf{r}}$ and $\boldsymbol{\sigma}_{\boldsymbol{\omega}}$  as the drone's quality measure and evaluate the performance of the FSO fronthaul link in terms of this measure.

In this paper, we assume that all position and orientation variables are independent from each other and follow Gaussian distributions, i.e., 
\begin{IEEEeqnarray}{lll} \label{Eq:Gaussian}
\mathbf{r} = \big(\mu_x+\epsilon_x,
\mu_y+\epsilon_y,\mu_z+\epsilon_z\big)
\qquad\IEEEyesnumber\IEEEyessubnumber \\
\boldsymbol{\omega}=  \big(\mu_{\theta}+\epsilon_{\theta}, \mu_{\phi}+\epsilon_{\phi}\big),
\IEEEyessubnumber
\end{IEEEeqnarray}
where $\epsilon_s\sim\mathcal{N}(0,\sigma^2_s)$ denotes a zero-mean normal RV with variance $\sigma^2_{s},\,s\in\{x,y,z,\theta,\phi\}$. We emphasize that the exact distributions of $\mathbf{r}$ and $\boldsymbol{\omega}$ have to be found via experimental measurements. Nevertheless, the adopted Gaussian model is a reasonble choice as it takes into account the first and second order moments of the RVs. Moreover, our assumption is inline with the Gaussian assumption made for derivation of the statistical model for the geometric loss due to building sway \cite{Steve_pointing_error,Alouini_Pointing}. Substituting normal RVs $\mathbf{r}$ and $\boldsymbol{\omega}$ into (\ref{Eq:PowerPhotoDetector}) and (\ref{Eq:hg_mean}), we obtain RVs $h_g(\mathbf{r},\boldsymbol{\omega})$ and $\tilde{h}_g(\mathbf{r},\boldsymbol{\omega})$, respectively. 

Note that in (\ref{Eq:hg_mean}), $A_0$, $k_{\mathrm{mean}}$, and $u$ are RVs since $A_0$ and $k_{\mathrm{mean}}$ depend on RV $\boldsymbol{\omega}$ and $u$ depends on both RVs $\mathbf{r}$ and $\boldsymbol{\omega}$. Nevertheless, we observed from our simulations that the variances of $A_0$ and $k_{\mathrm{mean}}$ are several orders of magnitude smaller than the variance of $u$. The reason for this behaviour is that a small variation in $\boldsymbol{\omega}$, e.g., on the order of mrad, has a significant effect on $u=\sqrt{f_y^2+f_z^2}$ since the impact of this variation on $f_y$ and $f_z$ in (\ref{Eq:FootPrint_Center}) is scaled by $r_x$ which typically has a very large value, i.e., on the order of several hundred meters. On the other hand, the impact of variations in $\boldsymbol{\omega}$ is not scaled by $r_x$. Therefore,  the main reason for the fluctuation of the geometric loss is the variation of the center of the beam footprint on the PD plane, i.e., $u$. Hence, in the following, we assume that the values of $A_0$ and $k_{\mathrm{mean}}$  are approximately constant and obtained based on the mean position and mean orientation of the drone, i.e., $\boldsymbol{\mu}_{\mathbf{r}}$ and $\boldsymbol{\mu}_{\boldsymbol{\omega}}$. Under this assumption, we determine the distribution of $u$ in the following theorem.

\begin{theo}\label{Theo:PDF_r}
Assuming $\sigma_s\to 0,\,s\in\{x,y,z,\theta,\phi\}$, the distance between the center of the beam and the center of the PD, $r$, follows a Hoyt (Nakagami-q) distribution $u\sim\mathcal{H}(q,\Omega)$ with parameters $q=\sqrt{\frac{\min\{\lambda_1,\lambda_2\}}{\max\{\lambda_1,\lambda_2\}}}$ and $\Omega=\lambda_1+\lambda_2$. Moreover,  $\lambda_1$ and $\lambda_2$ are the eigenvalues of matrix $\boldsymbol{\Sigma}$ which is given by
\begin{IEEEeqnarray}{lll} \label{Eq:Cov}
\boldsymbol{\Sigma} =
\begin{bmatrix}
\sigma^2_y+c_1^2\sigma^2_x+c_2^2\sigma^2_{\theta} 
& c_1c_5\sigma^2_x + c_2c_4\sigma^2_{\theta}\\
c_1c_5\sigma^2_x + c_2c_4\sigma^2_{\theta}
&  \sigma^2_z+c_3^2\sigma^2_{\phi}+c_4^2\sigma^2_{\theta}+c_5^2\sigma^2_x
\end{bmatrix},\quad
\end{IEEEeqnarray}
where $c_1=-\tan(\mu_{\theta})$, $c_2=-\frac{\mu_x}{\cos^2(\mu_{\theta})}$, $c_3=\frac{\mu_x}{\sin^2(\mu_{\phi})\cos(\mu_{\theta})}$, $c_4=-\frac{\mu_x\cot(\mu_{\phi})\tan(\mu_{\theta})}{\cos(\mu_{\theta})}$, and $c_5=-\frac{\cot(\mu_{\phi})}{\cos(\mu_{\theta})}$ are constants.
\end{theo}
\begin{IEEEproof}
Please refer to Appendix~\ref{App:Theo_PDF_r}.
\end{IEEEproof}

Based on the distribution of $u$ in Theorem~\ref{Theo:PDF_r}, the probability density function (PDF) of $\tilde{h}_g(\mathbf{r},\boldsymbol{\omega})$ in (\ref{Eq:hg_mean}), denoted by $f_{\tilde{h}_g}(x)$, is obtained as
\begin{IEEEeqnarray}{lll} \label{Eq:PDF}
f_{\tilde{h}_g}(x) = & \frac{\varpi }{A_0}
\left(\frac{x}{A_0}\right)^{\frac{(1+q^2)\varpi}{2q}-1} \nonumber \\
& \times I_0\left(-\frac{(1-q^2)\varpi}{2q}\ln\left(\frac{x}{A_0}\right)\right), \quad 0\leq x \leq A_0,\quad 
\quad
\end{IEEEeqnarray}
where $\varpi = \frac{(1+q^2)k_{\mathrm{mean}}w^2(L)}{4q\Omega}$ is a constant and $I_0(\cdot)$ is the zero-order modified Bessel function of the first kind.

Now, we consider the special case where the average position and orientation of the drone correspond to a beam which is orthogonal w.r.t. the PD plane. In other words, we have $\mu_y=\mu_z=0$, $\mu_{\theta}=\pi$, and $\mu_{\phi}=\pi/2$. This leads to the following simplified matrix $\boldsymbol{\Sigma}$
\begin{IEEEeqnarray}{lll} \label{Eq:Cov_Rayleigh}
\boldsymbol{\Sigma} =
\begin{bmatrix}
\sigma^2_y+\mu_{x}^2\sigma^2_{\theta}  & 0\\
0&  \sigma^2_z+\mu_x^2\sigma^2_{\phi}
\end{bmatrix},\quad
\end{IEEEeqnarray}
which has eigenvalues $\lambda_1=\sigma^2_y+\mu_{x}^2\sigma^2_{\theta}$ and $\lambda_2=\sigma^2_z+\mu_x^2\sigma^2_{\phi}$. Hereby, assuming $\sigma^2_y=\sigma^2_z\triangleq\sigma^2_{p}$ and $\sigma^2_{\theta} =\sigma^2_{\phi}\triangleq\sigma^2_{o}$, RV $u$ follows a Rayleigh distribution \cite{Steve_pointing_error} and $\tilde{h}_g(\mathbf{r},\boldsymbol{\omega})$ follows the following distribution
\begin{IEEEeqnarray}{lll} \label{Eq:PDF_Rayleigh}
f_{\tilde{h}_g}(x) = & \frac{\varrho}{A_0}
\left(\frac{x}{A_0}\right)^{\varrho-1}, \quad 0\leq x \leq A_0,\quad 
\end{IEEEeqnarray}
where $\varrho=\frac{k_{\mathrm{mean}} w^2(L)}{4(\sigma^2_p+\mu^2_x\sigma^2_o)}$.

\begin{remk}\label{Remk:Special}
Note that as the values of $\lambda_1$ and $\lambda_2$ increase, the quality of the channel deteriorates since the probability of small values of  $\tilde{h}_g$ increases. 
The simplified matrix $\boldsymbol{\Sigma}$ in (\ref{Eq:Cov_Rayleigh}) provides the important insight that the geometric loss is much more sensitive to the variance of the orientation $\sigma^2_{a}$ than to the variance of the position $\sigma^2_{p}$ since the variance of the orientation $\sigma^2_{a}$ is scaled by the mean distance between the drone and the CU, i.e., $\|\mathbf{r}\|=\mu_x$. 
\end{remk}

\section{Simulation Results}\label{Sec:Sim}
\begin{figure}
  \centering
\resizebox{1\linewidth}{!}{\psfragfig{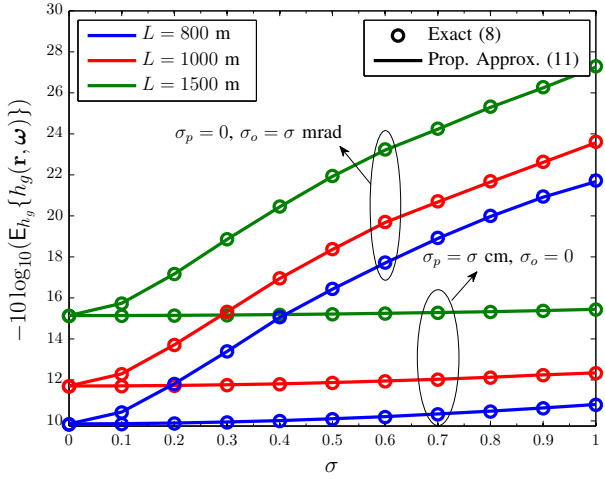}} 
\vspace{-8mm}
\caption{Average geometric loss (in dB) vs. drone's stability parameter $\sigma$ for $\alpha=\pi/8$, $\beta=5\pi/8$,  and  $L\in\{800,1000,1500\}$~m.  } \vspace{-2mm}
\label{Fig:Stat}
\end{figure}

\begin{figure}
  \centering
\resizebox{1\linewidth}{!}{\psfragfig{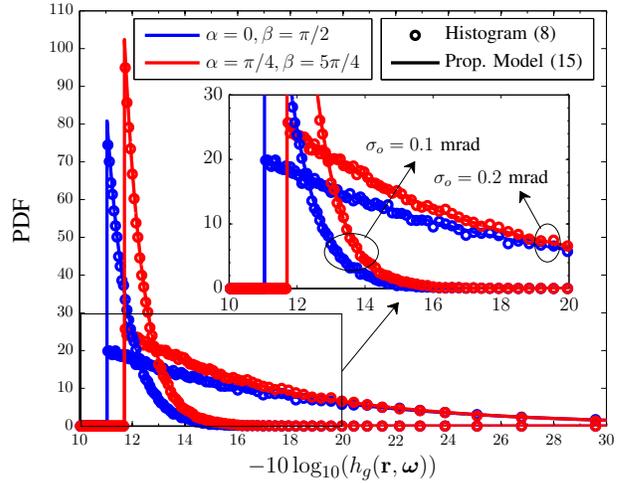}} 
\vspace{-8mm}
\caption{PDF of geometric loss for $\sigma_p=0$, $\sigma_o\in\{0.1,0.2\}$ mrad, and $(\alpha,\beta)\in\{(0,\pi/2),(\pi/8,5\pi/8)\}$.  } \vspace{-2mm}
\label{Fig:PDF}
\end{figure}

Unless stated otherwise, the default values of the parameters used for simulation are given by $C_n^2=10^{-14}$~m$^{2/3}$, $\lambda=1550$~nm, $a=10$~cm, $L=1$~km, $w_0=1$~mm \cite{Steve_pointing_error,Alouini_Pointing}. Moreover, we obtained the simulation results reported in Figs.~\ref{Fig:Stat} and \ref{Fig:PDF} based on $10^5$ realizations of RVs $\mathbf{r}$ and $\boldsymbol{\omega}$.
To better quantify the non-orthogonality of the beam w.r.t. the PD plane, we express the mean position of the drone, $\boldsymbol{\mu}_{\mathbf{r}}$, in spherical coordinates as $(R,\alpha,\beta)$, i.e., $r_x=R\sin(\beta)\cos(\alpha), r_y=R\sin(\beta)\sin(\alpha)$, and $r_z  = R\cos(\beta)$. Recall that for given $\boldsymbol{\mu}_{\mathbf{r}}$, $\boldsymbol{\mu}_{\boldsymbol{\omega}}$ is obtained from (\ref{Eq:Angles}). Moreover, we assume identical standard deviations for the position variables, i.e., $\sigma_x=\sigma_y=\sigma_z= \sigma_p$, and identical standard deviations for the orientation variables, i.e., $\sigma_{\theta}=\sigma_{\phi}= \sigma_o$. 

\begin{figure}
  \centering
\resizebox{1\linewidth}{!}{\psfragfig{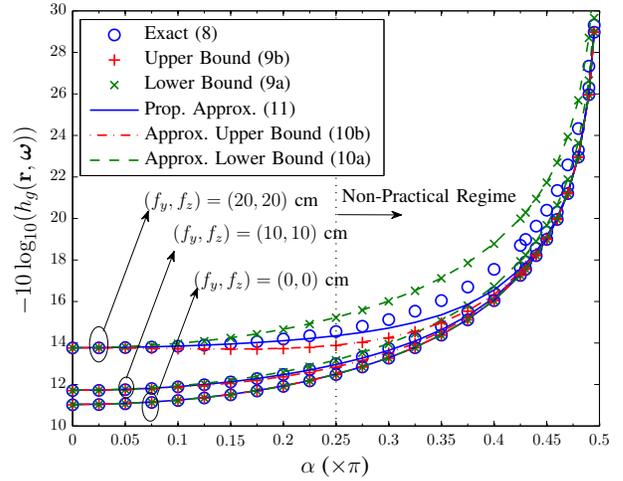}} 
\vspace{-8mm}
\caption{Geometric loss (in dB) vs. $\alpha$ for $\beta=\pi/2$, and $\sigma_p=\sigma_o=0$.  } \vspace{-2mm}
\label{Fig:Bound}
\end{figure}

First, we study the effect of the non-orthogonality of the beam w.r.t. the PD plane for the deterministic geometric loss and investigate the accuracy of the bounds proposed in Theorem~\ref{Theo:Power}, their approximations in (\ref{Eq:upper_lower_cal}), and the proposed approximation in (\ref{Eq:hg_mean}).  In Fig.~\ref{Fig:Bound}, we show the deterministic loss due to the geometric loss in decibel (dB), i.e., $-10\log_{10}(h_g(\mathbf{r},\boldsymbol{\omega}))$, vs. $\alpha$ for $\beta=\pi/2$, $\sigma_p=\sigma_o=0$, and different values for the center of the beam footprint $(f_y,f_z)$.  At $\alpha=0$, we have the special case of an orthogonal beam w.r.t. the PD plane where the loss for $(f_y,f_z)= (0,0)$ is due to the geometric loss and the loss for $(f_y,f_z)\neq (0,0)$ is due to both geometric loss and misalignment loss, cf. Footnote~\ref{Ftn:Loss}. We observe from Fig.~\ref{Fig:Bound} that as $\alpha$ increases, the loss increases. Note that although the beam line of the laser may not be orthogonal w.r.t. the PD plane, i.e., $\alpha\neq0$, $\alpha$ will be small in practice, i.e., $|\alpha|\ll \pi/2$. From Fig.~\ref{Fig:Bound}, we observe that for $\alpha < \pi/4$, the loss due to the non-orthogonality of the  beam is small (less than $1.5$~dB). Finally, Fig.~\ref{Fig:Bound} reveals that the proposed bounds and  approximations are accurate for practical values of $\alpha$, i.e., $\alpha<\pi/4$.

Next, we study the effect of random fluctuations of the position and orientation of the drone on the \textit{average} geometric loss, i.e., $\mathsf{E}_{h_g}\{h_g(\mathbf{r},\boldsymbol{\omega})\}$ where $\mathsf{E}_{x}\{\cdot\}$ denotes expectation w.r.t. RV $x$. Here, $\mathsf{E}_{h_g}\{h_g(\mathbf{r},\boldsymbol{\omega})\}$ is evaluated via simulation. In Fig.~\ref{Fig:Stat}, we show the average  geometric loss in dB vs. $\sigma$ for $\alpha=\pi/8$, $\beta=5\pi/8$, and different distances between the drone and the CU $L\in\{800,1000,1500\}$~m. Here, $\sigma$ is the standard deviation of the position or orientation, i.e., $(\sigma_p,\sigma_o)=(0,\sigma\,(\mathrm{cm}))$ or  $(\sigma_p,\sigma_o)=(\sigma\,(\mathrm{mrad}),0)$. 
We observe from Fig.~\ref{Fig:Stat} that the proposed approximation in (\ref{Eq:hg_mean}) closely approaches the exact value of the geometric loss in (\ref{Eq:PowerPhotoDetector}). Also, Fig.~\ref{Fig:Stat} reveals that the power loss due to orientation fluctuations on the order of mrad is much more severe than that due to position fluctuations on the order of cm. Furthermore, as the distance between the drone and the CU increases, the average geometric loss increases, too.

 
 In Fig.~\ref{Fig:PDF}, the PDF of the geometric loss is plotted for $\sigma_p=0$, $\sigma_o\in\{0.1,0.2\}$ mrad, and $(\alpha,\beta)\in\{(0,\pi/2),(\pi/8,5\pi/8)\}$. As can be observed from Fig.~\ref{Fig:PDF}, the  analytical statistical model proposed in (\ref{Eq:PDF}) is in perfect agreement with the histogram of (\ref{Eq:PowerPhotoDetector}). Note that the PDFs for the case where the beam is orthogonal w.r.t. the PD plane, i.e., $(\alpha,\beta)=(0,\pi/2)$, assume non-zero values at larger $h_g(\mathbf{r},\boldsymbol{\omega})$ (smaller $-10\log_{10}(h_g(\mathbf{r},\boldsymbol{\omega}))$) compared to the case where the beam is non-orthogonal w.r.t. the PD plane, i.e.,  $(\alpha,\beta)=(\pi/4,5\pi/4)$, cf. (\ref{Eq:PDF}). Moreover, as the standard deviation $\sigma_o$ increases, the probability of larger geometric losses increases and hence, the corresponding PDFs become more heavy tailed.

\section{Conclusions}

In this paper, we considered the modelling of a drone-based FSO fronthaul channel by quantifying the geometric loss caused by random fluctuations of the position and the orientation of the drone. We derived upper and lower bounds, corresponding approximate expressions, and a closed-form statistical model for the geometric loss. Furthermore, we validated our derivations via simulations and quantified the impact of the drone's instability on the quality of the FSO channel using the developed model for the geometric loss. In future work, the proposed analytical model can be exploited for performance analysis of the considered drone-based communication system in terms of e.g. the outage probability, average bit/symbol error rate, and average achievable rate.

\appendices
\section{Proof of Lemma~\ref{Lem:PowerDensity}}\label{App:Lem_PowerDensity}

Note that $I(y,z)\mathrm{d}y\mathrm{d}z$ determines the fraction of power collected in the infinitesimally small area $\mathrm{d}y\mathrm{d}z$, i.e.,  $\mathrm{d}y\to 0$ and $\mathrm{d}z \to 0$, around the point $(0,y,z)$. Recall that the power density in any perpendicular plane is given by (\ref{Eq:PowerOrthogonal}). To exploit this knowledge, we use the fact that any point  $(0,y,z)$ in the PD plane is also located in another plane which is perpendicular to the beam line. Therefore, the power $I(y,z)\mathrm{d}y\mathrm{d}z$ can be obtained as $I(y,z)\mathrm{d}y\mathrm{d}z=I^{\mathrm{orth}}(L,l)\sin(\psi)\mathrm{d}y\mathrm{d}z$, where $\psi=\sin^{-1}(\sin(\phi) \cos(\theta))$. Next, we find $L$ and $l$.
 The distance $l$ is the distance between point $(0,y,z)$ and the beam line in (\ref{Eq:BeamLine}). In general, the distance between a point $\mathbf{p}$ and a line specified by direction vector $\mathbf{u}$ and a given point $\mathbf{q}$ on the line can be obtained as
\begin{IEEEeqnarray}{lll} \label{Eq:DistanceDefine}
l = \frac{\Vert (\mathbf{p}-\mathbf{q})\times \mathbf{u}\Vert}{\Vert \mathbf{u}\Vert },
\end{IEEEeqnarray}
where $\times$ denotes the cross product between two vectors. For the problem at hand, we choose $\mathbf{p}=(0,y,z)$, $\mathbf{u}=\mathbf{d}$, and $\mathbf{q}=\mathbf{f}$, which leads to 
\begin{IEEEeqnarray}{lll} \label{Eq:Distance}
l = \Vert ((0,y,z)-\mathbf{f})\times \mathbf{d} \Vert =\\
\Big\Vert\big[\tilde{y}\cos(\phi)-\tilde{z}\sin(\phi)\sin(\theta),
\tilde{z}\sin(\phi)\cos(\theta),\tilde{y}\sin{\phi}\cos(\theta)\big]\Big\Vert \nonumber\\
=\rho_y\tilde{y}+\rho_z\tilde{z}+2\rho_{yz}\tilde{y}\tilde{z},\nonumber
\end{IEEEeqnarray}
where we exploited the fact that $\Vert \mathbf{d}\Vert=1$ and introduced $\tilde{y} = y-f_y$ and $\tilde{z} =z-f_z$ where $\rho_y$, $\rho_z$, and $\rho_{yz}$ are given in Lemma~\ref{Lem:PowerDensity}. Moreover, the distance between the perpendicular plane and the laser source can be bounded as
\begin{IEEEeqnarray}{lll} 
\Vert \mathbf{r} - \mathbf{f} \Vert -  \sqrt{\tilde{y}^2+\tilde{z}^2} \leq L \leq \Vert \mathbf{r} - \mathbf{f} \Vert+ \sqrt{\tilde{y}^2+\tilde{z}^2},
\end{IEEEeqnarray}
where the extreme cases occur if the the beam line is parallel to plane $x=0$. In particular, we can safely assume that $\Vert \mathbf{r} - \mathbf{f} \Vert \pm  \sqrt{\tilde{y}^2+\tilde{z}^2} \approx \Vert \mathbf{r} \Vert$ holds since the distance between the drone and the CU is much larger than $\Vert \mathbf{f} \Vert$ and $\sqrt{\tilde{y}^2+\tilde{z}^2}$. Substituting these results in  (\ref{Eq:PowerOrthogonal}) leads to (\ref{Eq:Intensity}) and completes the proof.

\section{Proof of Theorem~\ref{Theo:Power}}\label{App:Integral}
Note that in the $(y,z)$ plane, the contours of power density $I(y,z)=\bar{I}$ form ellipsoids given by 
\begin{IEEEeqnarray}{lll} \label{Eq:ellipse}
\rho_y(y-f_y)^2+2\rho_{yz}(y-f_y)(z-f_z)+\rho_z(z-f_z)^2=d,\qquad
\end{IEEEeqnarray}
where $d=\frac{w^2(L)}{2}\log\left(\frac{2\sin(\psi)}{\pi w^2(L)\bar{I}}\right)$. These ellipsoids are centered at point $(f_y,f_z)$ and rotated by angle $ \gamma=\frac{1}{2}\tan^{-1}\big(\frac{2 \rho_{yz}}{\rho_{y} -\rho_{z}}\big)$ counterclockwise, and have minor and major axis lengths of $\sqrt{\rho_{\min}/d}$ and $\sqrt{\rho_{\max}/d}$, respectively, where $\rho_{\min}=\frac{2}{\rho_y+\rho_z+\sqrt{(\rho_y-\rho_z)^2+4\rho_{zy}^2}}$ and $\rho_{\max}=\frac{2}{\rho_y+\rho_z-\sqrt{(\rho_y-\rho_z)^2+4\rho_{zy}^2}}$.

In order to obtain the lower and upper bounds of $h_g(\mathbf{r},\boldsymbol{\omega})$, we substitute the contour in (\ref{Eq:ellipse}) by two rotated elliptic contours which have the same axis lengths $\rho_{\min}$ and $\rho_{\max}$; however, their main axes are either parallel or perpendicular to the line connecting $(f_y,f_z)$ and the origin, see Fig.~\ref{Fig:Contour}. Moreover, without loss of generality, we can define a new coordinate system by rotating the $y$ and $z$ axes by angle $\tau=\tan^{-1}(\frac{f_z}{f_y})$ such that the center of the ellipsoid in (\ref{Eq:ellipse}) lies on the rotated $y$ axis, i.e., the center becomes $\Big(\big(\sqrt{f_y^2+f_z^2}\big),0\Big)$ in the new coordinate system. Note that the circular PD has the same description in the new and the old coordinate systems.  This leads to lower and upper bounds $h_g^{\mathrm{low}}(\mathbf{r},\boldsymbol{\omega})$ and $h_g^{\mathrm{upp}}(\mathbf{r},\boldsymbol{\omega})$, respectively, as given in Theorem~1. This completes the proof.

\section{Proof of Theorem~\ref{Theo:PDF_r}}\label{App:Theo_PDF_r}
We use the Taylor series expansions of $\tan(\theta)$, $\cot(\phi)$, and $\frac{1}{\cos(\theta)}$  in the expressions for $f_y$ and $f_z$, i.e.,
\begin{IEEEeqnarray}{lll} \label{Eq:Taylor}
\underset{\theta\to\mu_{\theta}}{\lim} \tan(\theta) &= \tan(\mu_{\theta})+(\theta-\mu_{\theta})\frac{1}{\cos^2(\mu_{\theta})}\qquad\IEEEyesnumber\IEEEyessubnumber\\
\underset{\phi\to\mu_{\phi}}{\lim} \cot(\phi)
&= \cot(\mu_{\phi})-(\phi-\mu_{\phi})\frac{1}{\sin^2(\mu_{\phi})}\qquad\IEEEyessubnumber\\
\underset{\theta\to\mu_{\theta}}{\lim} \frac{1}{\cos(\theta)}
&= \frac{1}{\cos(\mu_{\theta})}+(\theta-\mu_{\theta})\frac{\tan(\mu_{\theta})}{\cos(\mu_{\theta})}.\qquad\IEEEyessubnumber
\end{IEEEeqnarray}
We note that from (\ref{Eq:Gaussian}), we have $\epsilon_{\theta}= \theta-\mu_{\theta}$ and $\epsilon_{\phi}=\phi-\mu_{\phi}$. Substituting (\ref{Eq:Taylor}) into (\ref{Eq:FootPrint_Center}) and simplifying the results assuming perfect beam tracking using (\ref{Eq:Angles}), we obtain
\begin{IEEEeqnarray}{lll} \label{Eq:fyfz_norm}
\underset{s\to\mu_{s},\,\forall s\in\{x,y,z,\theta,\phi\}}{\lim}  f_y = \epsilon_y+c_1\epsilon_x+c_2\epsilon_{\theta}\qquad\IEEEyesnumber\IEEEyessubnumber\\
\underset{s\to\mu_{s},\,\forall s\in\{x,y,z,\theta,\phi\}}{\lim} f_z = \epsilon_z+c_3\epsilon_{\phi}+c_4\epsilon_{\theta}+c_5\epsilon_x,\qquad\IEEEyessubnumber
\end{IEEEeqnarray}
where constants $c_1$, $c_2$, $c_3$, $c_4$, and $c_5$ are given in Theorem~\ref{Theo:PDF_r}. To obtain (\ref{Eq:fyfz_norm}), we dropped the terms with orders higher than one, e.g., $\epsilon_{\theta}\epsilon_{\phi}$.
Since $f_y$ and $f_z$ in  (\ref{Eq:fyfz_norm}) are sums of Gaussian RVs, they are Gaussian distributed, too. However, $f_y$ and $f_z$ are correlated since $\epsilon_{x}$ and $\epsilon_{\theta}$ appear in both of them. The joint distribution of $f_y$ and $f_z$ is a bivariate Gaussian distribution $(f_y,f_z)\triangleq \bar{\mathbf{f}}\sim\mathcal{N}(\mathbf{0},\boldsymbol{\Sigma})$ where $\mathbf{0}=(0,0)$ and $\boldsymbol{\Sigma}$ is given in (\ref{Eq:Cov}). Let $\boldsymbol{\Sigma}=\mathbf{U}\boldsymbol{\Lambda}\mathbf{U}^\mathsf{T}$ be the eigenvalue decomposition of $\boldsymbol{\Sigma}$ where $\boldsymbol{\Lambda}$ is a diagonal matrix with elements $\lambda_1$ and $\lambda_2$, $\mathbf{U}$ is a unitary matrix, i.e., $\mathbf{U}^\mathsf{T}\mathbf{U}=\mathbf{I}$, where $\mathbf{I}$ is the identity matrix and $(\cdot)^{\mathsf{T}}$ denotes the  transpose operation. Using these definitions, it is easy to show that $\bar{\mathbf{f}}\sim\mathbf{g}\mathbf{U}^\mathsf{T}$ where $\mathbf{g}=(g_y,g_z)\sim\mathcal{N}(\mathbf{0},\boldsymbol{\Lambda})$. Now, we can express $u$ in terms of $\mathbf{g}$ as 
\begin{IEEEeqnarray}{lll} 
u =\sqrt{f_y^2+f_z^2}=\sqrt{\bar{\mathbf{f}}\bar{\mathbf{f}}^\mathsf{T}} \sim \sqrt{\mathbf{g}\mathbf{U}^\mathsf{T}\mathbf{U} \mathbf{g}^\mathsf{T}}
= \sqrt{g_y^2+g_z^2}. \quad
\end{IEEEeqnarray}
Since $g_y$ and $g_z$ are independent zero-mean RVs with non-identical variances,  $u$ follows a Hoyt (Nakagami-q) distribution $u\sim\mathcal{H}(q,\Omega)$, where $q=\sqrt{\frac{\min\{\lambda_1,\lambda_2\}}{\max\{\lambda_1,\lambda_2\}}}$ and $\Omega=\lambda_1+\lambda_2$ \cite{Nakagami_Hoyt}. This completes the proof.

\bibliographystyle{IEEEtran}
\bibliography{My_Citation_19-10-2017}

\end{document}